\documentclass[a4paper,11pt]{article}
\usepackage{jheppub} % for details on the use of the package, please see the JINST-author-manual
\usepackage{physics}
\usepackage{amsmath}
\usepackage{amssymb}
\usepackage{graphicx}
\usepackage{overpic}
\usepackage{authblk}
\usepackage{epsfig}
\usepackage{epstopdf}
\usepackage{colortbl}
\usepackage{subcaption}
\usepackage{mathrsfs}

\usepackage{rotating}
\usepackage{bm}% bold math
\usepackage{dcolumn}% Align table columns on decimal point
\usepackage{multirow}
\usepackage{makecell}
\usepackage{indentfirst}
\usepackage{booktabs}
\usepackage{arydshln}
\usepackage{xcolor}

\newcommand{\BESIIIorcid}[1]{\href{https://orcid.org/#1}{\hspace*{0.1em}\raisebox{-0.45ex}{\includegraphics[width=1em]{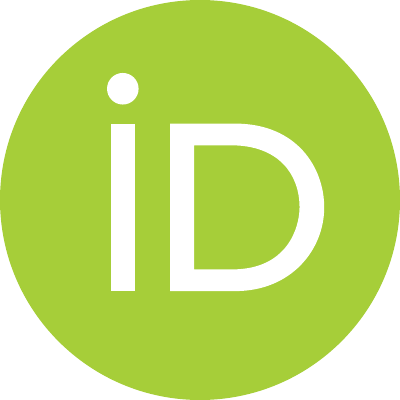}}}} 

\title{\boldmath Search for the charged lepton flavour violating decay $\eta'\to e\mu$}

\collaboration{
BESIII Collaboration\hyperref[author]{*}\\
\includegraphics[height=15mm]{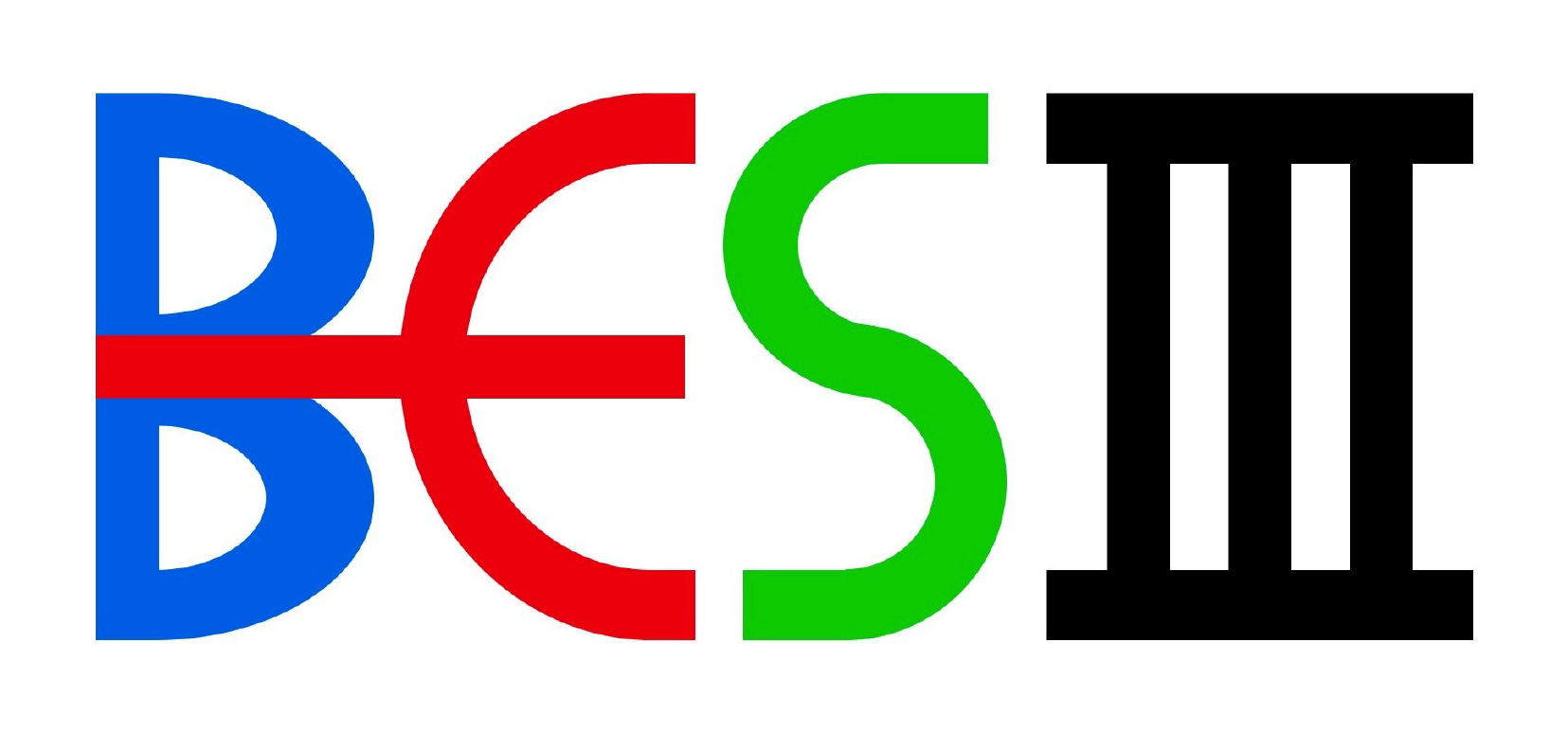} 
}

\emailAdd{besiii-publications@ihep.ac.cn}

\abstract{
Based on $(8998\pm40)\times10^6$ $J/\psi$ events collected in $e^+e^-$ collisions at $\sqrt{s} = 3.097~{\rm GeV}$ with the BESIII detector, we present a search for the charged lepton flavour violating decay $\eta'\to e\mu$ with $J/\psi\to\gamma\eta'$. No significant signal is observed, and an upper limit on its decay branching fraction is set to be $6.3\times10^{-7}$ at the 90\% confidence level, improving the previous best result by nearly three orders of magnitude.
}


\begin{document}
\maketitle
\flushbottom

%%====================================================

\section{Introduction}
\label{sec:intro}

The observation of neutrino oscillations established that lepton flavour is not an exact symmetry in nature, implying that charged lepton flavour violating (CLFV) processes can occur through neutrino mixing loops in the Standard Model extended with neutrino masses. However, the predicted branching fractions are extremely suppressed due to the small neutrino masses~\cite{bib:1-1}.
Any observation of a CLFV process would be an unambiguous signature of new physics. Unlike other conservation laws in the Standard Model (SM), lepton flavour conservation is not associated with any underlying conserved current symmetry. As a result, many models beyond the SM naturally introduce CLFV transitions~\cite{bib:1-0}, such as the two Higgs doublet model with extra Yukawa couplings~\cite{bib:1-2.0}, the Minimal Supersymmetric extension of the SM~\cite{Brignole:2004ah}, the Grand Unified Theories~\cite{bib:1-3}, and the seesaw mechanisms~\cite{bib:1-4}. Some CLFV processes are predicted to occur at rates close to the current experimental limits in certain parameter-dependent models~\cite{bib:1-5,bib:1-6}.

\begin{figure}[!ht]
    \centering
    \begin{minipage}[t]{0.32\textwidth}
        \centering
        \includegraphics[width=0.7\textwidth]{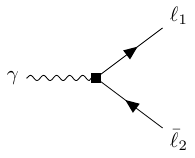}
        \subcaption{Dipole interaction.}
        \label{fg:0.1}
    \end{minipage}
    \begin{minipage}[t]{0.32\textwidth}
        \centering
        \includegraphics[width=0.7\textwidth]{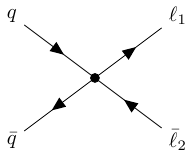}
        \subcaption{Four-fermion interaction.}
        \label{fg:0.2}
    \end{minipage}
    \begin{minipage}[t]{0.32\textwidth}
        \centering
        \includegraphics[width=0.7\textwidth]{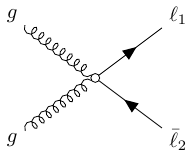}
        \subcaption{Gluon-fermion interaction.}
        \label{fg:0.3}
    \end{minipage}
    \caption{The Feynman diagrams of three kinds of interaction introduced in the effective field theory~\cite{bib:1-7}. The $\eta'\to e\mu$ process is related to the last two kinds of interaction.}
    \label{fg:0}
\end{figure}

To encode various theoretical models for comparison with experimental results, effective field theory is employed to describe CLFV processes. For CLFV decays of quarkonium, the effective Lagrangian can be divided into three parts~\cite{bib:1-7}: the dipole part, the part involving four-fermion interactions, and the gluonic part, as shown in Figure~\ref{fg:0}. The contributions of new physics mechanisms, represented by different higher-dimensional operators, are encoded in the Wilson coefficients. 
As reported in Reference~\cite{bib:1-7}, CLFV decays of quarkonium states with distinct quantum numbers and quark compositions are sensitive to different Wilson coefficients.
This makes it possible to select specific quarkonium states to probe the Wilson coefficients of interest.
 %Unless otherwise specified, quarkonium in this paper also includes light unflavoured mesons.

On the experimental side, considerable efforts have been devoted to searching for CLFV decays of leptons, mesons and bosons~\cite{bib:1-7.1}. Specifically, for quarkonium, searches have been conducted for pseudoscalar states ($\pi^{0}$~\cite{bib:1-8}, $\eta$~\cite{bib:1-9}, $\eta'$~\cite{bib:1-10}) and vector states ($\phi$~\cite{bib:1-11}, $J/\psi$~\cite{bib:1-12,bib:1-14}, $\psi(2S)$~\cite{bib:1-13}, $\Upsilon$~\cite{bib:1-15,bib:1-16,bib:1-17,bib:1-18}). Pseudoscalar states like $\eta'$ are sensitive to the axial coefficients ($C_{\rm AL/AR}^{q\ell_1\ell_2}$) and pseudoscalar coefficients ($C_{\rm PL/PR}^{q\ell_1\ell_2}$) in four-fermion interactions, as well as gluonic coefficients ($C_{\rm \tilde{G}L/\tilde{G}R}^{\ell_1\ell_2}$) in gluon-fermion interactions. The subscripts $L$ and $R$ specify the chirality of the corresponding lepton current.
For the $\eta'\to e\mu$ decay, The large $J/\psi$ sample available at BESIII enables a substantial improvement over the previous CLEO limit on the branching fraction (BF) of $4.7 \times 10^{-4}$ at 90\% confidence level (CL)~\cite{bib:1-10}.

In this analysis, we present a search for the CLFV process $\eta'\to e\mu$ via the decay $J/\psi\to\gamma\eta'$ utilizing $(8998\pm40)\times10^6$ $J/\psi$ events~\cite{bib:3-1} collected with the BESIII detector at the BEPCII collider. Based on the BF of $J/\psi\to\gamma\eta'$~\cite{pdg}, the number of produced $\eta'$ is estimated to be  $(47.5\pm0.6)\times10^6$. Throughout this paper, the notation $\eta'\to e\mu$ implicitly includes charge-conjugate final states. A stepwise unblinding analysis strategy is employed to avoid involuntary bias. The signal selection criteria are determined solely based on Monte Carlo (MC) simulation samples and independent continuum data samples, both of which are described in detail in Section~\ref{sec:bes}. The  analysis procedure includes the optimization of event selection criteria and the background study. A randomly selected subset of the total data sample is then used to validate the analysis method and optimize the event selection procedure. The refined analysis procedure is subsequently applied to the full data sample.

%%===================================================

\section{BESIII detector and data samples}
\label{sec:bes}

The BESIII detector~\cite{Ablikim:2009aa} records symmetric $e^+e^-$ collisions provided by the BEPCII storage ring~\cite{Yu:IPAC2016-TUYA01} in the center-of-mass (c.m.) energy range from 1.84 to 4.95 GeV, with a peak luminosity of $1.1 \times 10^{33}\;\text{cm}^{-2}\text{s}^{-1}$ 
achieved at $\sqrt{s} = 3.773\;\text{GeV}$. BESIII has collected large data samples in this energy region~\cite{Ablikim:2019hff}. The cylindrical core of the BESIII detector, covering 93\% of the full solid angle, consists of a helium-based multilayer drift chamber (MDC), a plastic scintillator time-of-flight system (TOF), and a CsI(Tl) electromagnetic calorimeter (EMC). All components are enclosed within a $1.0~\rm T$ superconducting solenoidal magnet. The solenoid is supported by an octagonal flux-return yoke with resistive plate muon counter (MUC) identification modules interleaved with steel. % The magnetic field was $0.9~\rm T$ in 2012, which affects 11\% of the total $J/\psi$ data. 
The charged-particle momentum resolution at $1~{\rm GeV}/c$ is $0.5\%$, and the energy loss $({\rm d}E/{\rm d}x)$ resolution is 6\% for electrons from Bhabha scattering. The EMC measures photon energies with a resolution of 2.5\% (5\%) at $1~\rm GeV$ in the barrel (end cap) region. The time resolution in the TOF barrel region is $68~\rm ps$, while that in the end cap region was $110~\rm ps$. The end cap TOF system was upgraded in 2015 using multigap resistive plate chamber technology, providing a time resolution of $60~\rm ps$, which benefits 98\% of the data used in this analysis~\cite{etof-1,etof-2,etof-3}.

MC simulated samples generated with a \textsc{GEANT4}-based~\cite{geant4} software package, which includes the geometric description~\cite{bib:2-1, bib:2-2, bib:2-3, bib:2-4} of the BESIII detector and the detector response, are used to determine detection efficiencies and estimate backgrounds. The simulation models the beam energy spread and initial state radiation in the $e^+e^-$ annihilation with the \textsc{KKMC} generator~\cite{ref:kkmc-1,ref:kkmc-2}. The inclusive MC sample includes the $J/\psi$ resonance  production incorporated in \textsc{KKMC}. All particle decays are simulated using \textsc{EVTGEN}~\cite{ref:evtgen-1, ref:evtgen-2}. The BFs are taken from the Particle Data Group (PDG)~\cite{pdg} when available, and estimated with \textsc{LUNDCHARM}~\cite{ref:lundcharm-1, ref:lundcharm-2} otherwise. Final-state radiation from charged final-state particles is incorporated using the \textsc{PHOTOS} package~\cite{photos2}.

To estimate the signal detection efficiency, a sample of 0.6 million signal MC events is generated. The process $J/\psi\to\gamma\eta'$ is simulated using the HELAMP model~\cite{ref:evtgen-1} which allows the simulation of any two-body decay by specifying the helicity amplitudes for the final-state particles. The subsequent process $\eta'\to e\mu$ is simulated using the PHSP model~\cite{ref:evtgen-1}, which assumes a uniform phase-space distribution for the decay. Additionally, an inclusive MC sample containing about 9 billion $J/\psi$ events, corresponding to the same statistics as the collected experimental data, is used to study the backgrounds from other $J/\psi$ meson decays.

In this paper, $(8998\pm40)\times10^6$  $J/\psi$ events~\cite{bib:3-1} collected at c.m. energies $(\sqrt{s})$ of $3.097~{\rm GeV}$ during 2009, 2018 and 2019 are analyzed, corresponding to an integrated luminosity $(\mathcal{L})$ of $2760~{\rm pb}^{-1}$. At this c.m. energy, no additional hadrons are produced together with the $J/\psi$ meson.
In addition, data samples collected at $\sqrt{s}=$ $3.080~{\rm GeV}$, $3.650~{\rm GeV}$, $3.682~{\rm GeV}$, $3.773~{\rm GeV}$, corresponding to the luminosities  of $166.3~{\rm pb}^{-1}$, $455~{\rm pb}^{-1}$, $404~{\rm pb}^{-1}$, $7911.9~{\rm pb}^{-1}$~\cite{bib:3-2.1,bib:3-2.2}, are used to study the continuum background under the $J/\psi$ resonance. In the following analysis, samples from these four energy points are scaled to the luminosity of the full $J/\psi$ data sample, weighted and summed into one continuum background sample. The scale factors are defined as $f_{i} = \frac{\mathcal{L}_{3.097}}{\mathcal{L}_{i}}\cdot\frac{s_{i}}{s_{3.097}}$, and the weights are defined as $w_{i} = \frac{\mathcal{L}_{i}}{\sum_{j}\mathcal{L}_{j}}$.

%%====================================================

\section{Event selection}
\label{sec:ana}

The signal process $J/\psi \to \gamma \eta'$ with $\eta' \to e\mu$ is characterized by final states consisting of one energetic photon and two oppositely charged tracks corresponding to an electron and a muon. To reconstruct this topology, the selection requires exactly two charged tracks and employs particle identification (PID) to distinguish the lepton flavours.

Charged tracks detected in the MDC are required to fall within the polar angle range of $\abs{\cos\theta}<0.93$, where $\theta$ is defined with respect to the $z$-axis, the symmetry axis of the MDC. For all charged tracks, the distance of closest approach to the interaction point must be within $10~\rm cm$ along the $z$-axis and within $1~\rm cm$ in the transverse plane. Only events with exactly two selected tracks with opposite charges are retained for further analysis.

\begin{figure}[!ht]
    \centering
    \begin{minipage}[t]{0.49\textwidth}
        \includegraphics[width=\textwidth]{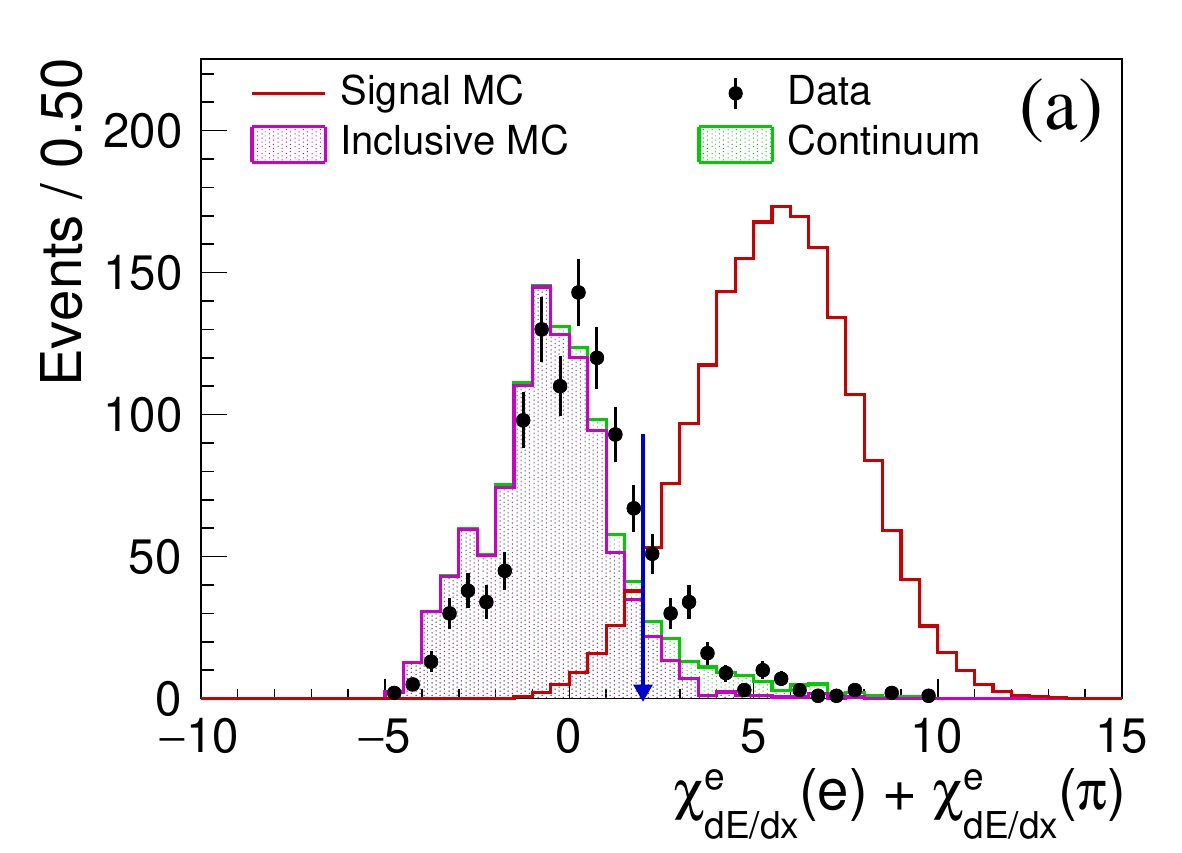}
    \end{minipage}
    \begin{minipage}[t]{0.49\textwidth}
        \includegraphics[width=\textwidth]{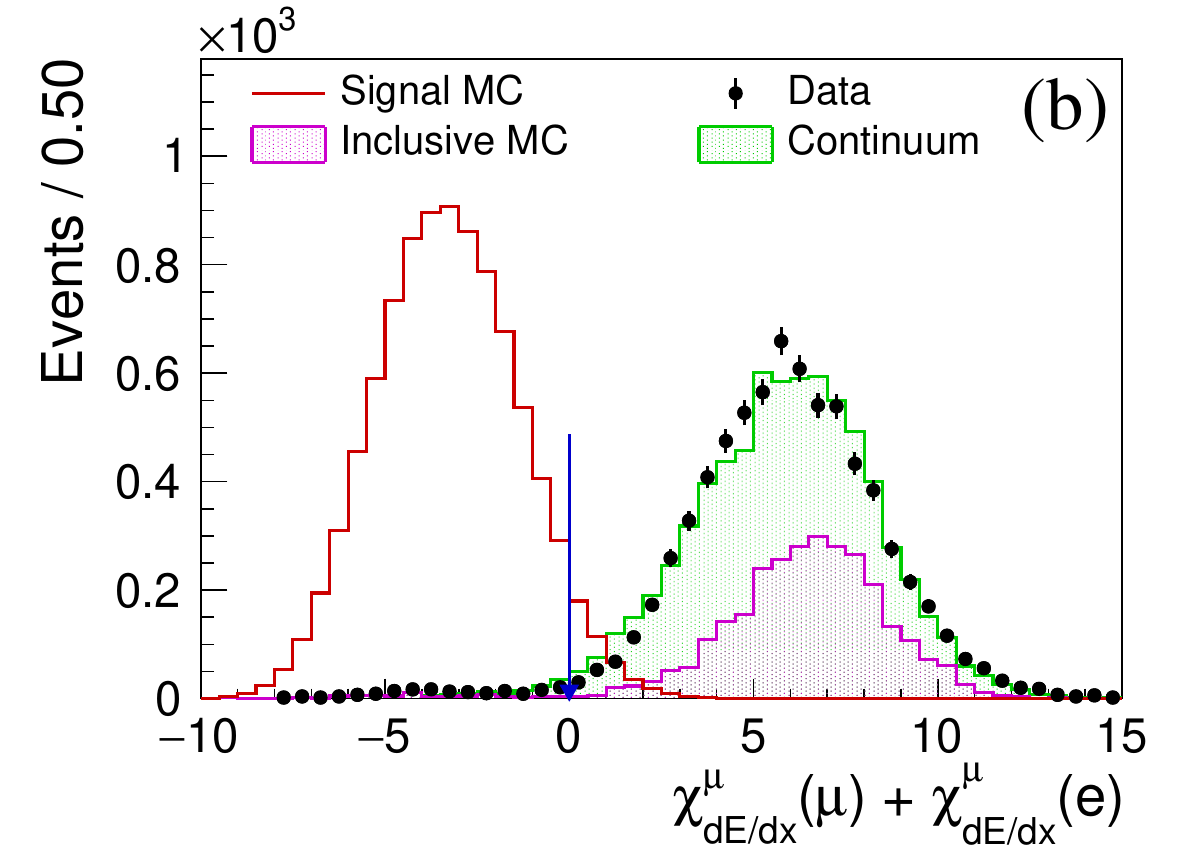}
    \end{minipage}
    \caption{The distribution of (a) $\chi_{\dd E/\dd x}^{e}(e)+\chi_{\dd E/\dd x}^{e}(\pi)$ and (b) $\chi_{\dd E/\dd x}^{\mu}(\mu)+\chi_{\dd E/\dd x}^{\mu}(e)$ of the accepted candidate events. The red histograms represent the signal MC sample and the black dots with error bars represent data sample. The green and magenta shaded histograms represent continuum samples and inclusive MC sample, respectively. The blue arrows indicate the selection criteria.}
    \label{fg:1}
\end{figure}

\begin{figure}[!ht]
    \centering
    \begin{minipage}[t]{0.49\textwidth}
        \includegraphics[width=0.9\textwidth]{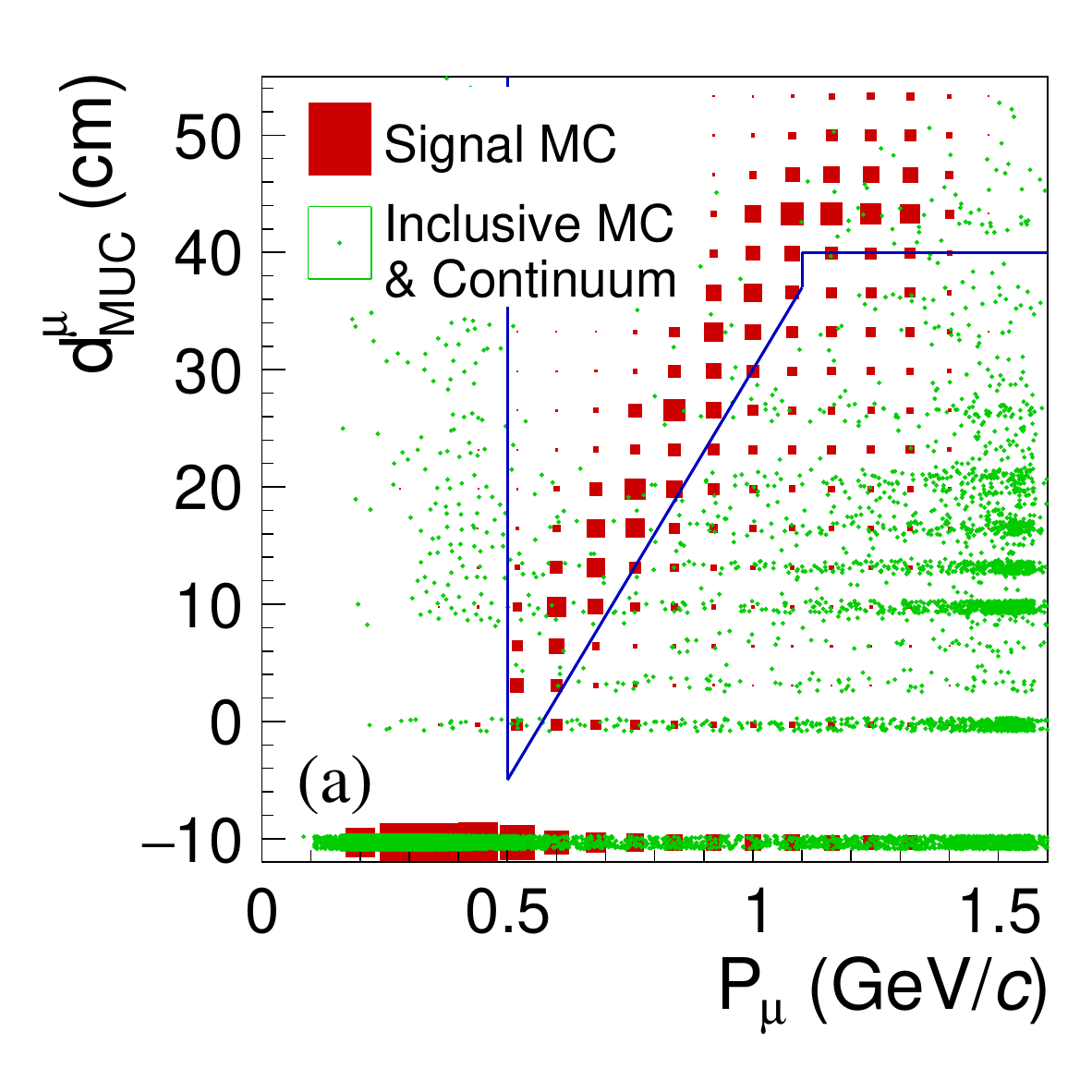}
        % \subcaption{Signal MC and background samples.}
    \end{minipage}
    \begin{minipage}[t]{0.49\textwidth}
        \includegraphics[width=0.9\textwidth]{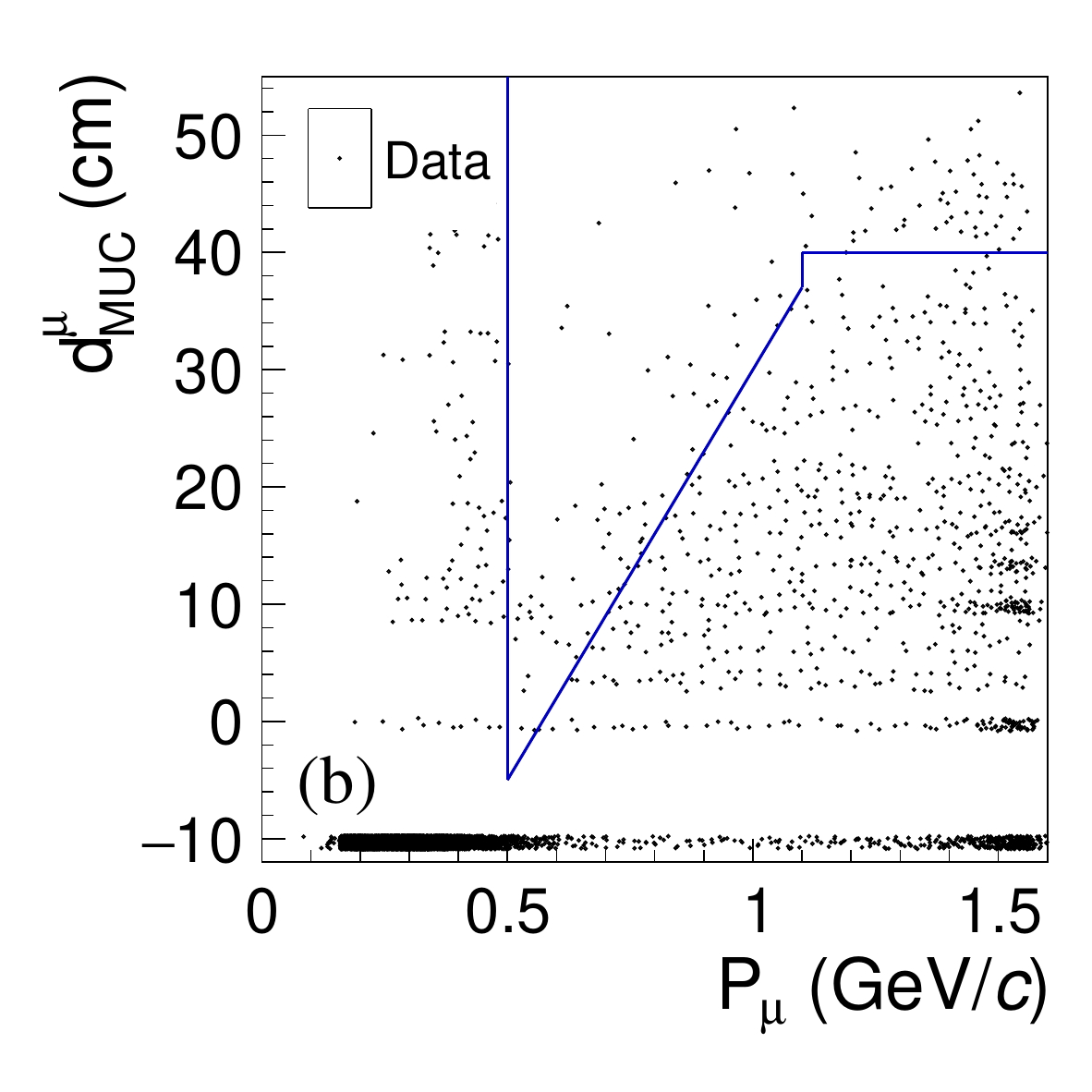}
        % \subcaption{Data.}
    \end{minipage}
    \caption{The two-dimensional distribution of $d_{\rm MUC}^{\mu}$ versus $P_{\mu}$ of the accepted candidate events. (a) The red boxes represent the signal MC sample, where the size of each box is proportional to the event yield in the corresponding bin, while the green dots show the background (inclusive MC sample and continuum samples). (b) The black dots represent the data sample. The events with $P_{\mu}>0.5~{\rm GeV}/c$ and lying above the solid blue line are identified as candidates with muon final state. The horizontal band at $-10$~cm corresponds to events, whose muon tracks lack reconstructed information in the MUC.}
    \label{fg:2}
\end{figure}

Since the dominant backgrounds arise from the events containing the final states $e^+e^-$ and $\pi^+\pi^-$, effective discrimination is required to separate the $e$-$\mu$, $e$-$\pi$, and $\mu$-$\pi$.
The identification probability $\mathcal{P}_i(j)$ of a candidate particle $i$ under particle hypothesis $j$ is determined using the measured ionization energy loss $(\dd E/\dd x)$ in the MDC, the flight time in the TOF, and the shower shape in the EMC. In this paragraph, $i = e, \mu$ and $j = e, \mu, \pi, K$.
Electron candidates must satisfy $\mathcal{P}_{e}(e)/\qty[\mathcal{P}_{e}(e)+\mathcal{P}_{e}(\pi)+\mathcal{P}_{e}(K)]>0.8$ and $\mathcal{P}_e(e)>0.001$.
In addition, the ratio of the deposited energy in the EMC to the momentum, $E_{\rm EMC}^{e}/P_e$, is required to be greater than 0.8 to further suppress the hadronic background.
To suppress the remaining $\pi^+\pi^-$ background, the standardized residual $\chi_{\dd E/\dd x}^i(j)$ is introduced, which is defined as the difference between the $\dd E/\dd x$ measured for candidate $i$ and that expected under hypothesis $j$, in units of the corresponding resolution. Figure~\ref{fg:1}(a) shows the distribution of $\chi_{\dd E/\dd x}^{e}(e)+\chi_{\dd E/\dd x}^{e}(\pi)$ for the electron candidates, and the requirement $\chi_{\dd E/\dd x}^{e}(e)+\chi_{\dd E/\dd x}^{e}(\pi)>2$ is applied. Meanwhile, the muon candidates must satisfy $\chi_{\dd E/\dd x}^{\mu}(\mu)+\chi_{\dd E/\dd x}^\mu(e)<0$ to suppress the $e^+e^-$ background, as shown in Figure~\ref{fg:1}(b).
The deposited energy of muon candidates in EMC, $E_{\rm EMC}^{\mu}$, is required to be between $0.1~{\rm GeV}$ and $0.3~{\rm GeV}$.
In addition, MUC detection information is applied to separate the muon from the pion.
Figure~\ref{fg:2} shows the momentum-dependent penetration depth of particles in the MUC.
A momentum-dependent cut on the penetration depth is implemented for muon identification.

Photon candidates are identified using the showers in the EMC. The deposited energy of each shower must exceed $25~\rm MeV$ in the barrel region $(\abs{\cos\theta} < 0.80)$ or $50~\rm MeV$ in the end cap region $(0.86 < \abs{\cos\theta}< 0.92)$. To suppress electronic noise and unrelated showers to the candidate event, the difference between the shower time and the event start time must lie within $[0, 700]~\rm ns$. In addition, to exclude showers associated with the charged tracks, the opening angle between the positions of an EMC shower and the closest charged track extrapolated to the EMC is required to be greater than $10^\circ$. 

Given that the signal photon $\gamma$ is produced in the two-body decay $J/\psi\to\gamma\eta'$ with the $J/\psi$ approximately at rest, the photon is nearly monochromatic with an energy of about $1.4~{\rm GeV}$, which is much higher than the typical photon energy from final-state radiation. The highest-energy photon is therefore identified as the photon from the $J/\psi$ decay.
A four-constraint (4C) kinematic fit is then performed on the $\gamma e\mu$ system to enforce overall energy-momentum conservation using the well-known initial four-momentum of the $e^+e^-$ collision. 
Events with $\chi_{\rm 4C}^2>100$ are discarded.

\begin{figure}[!ht]
    \centering
    \includegraphics[width=0.5\linewidth]{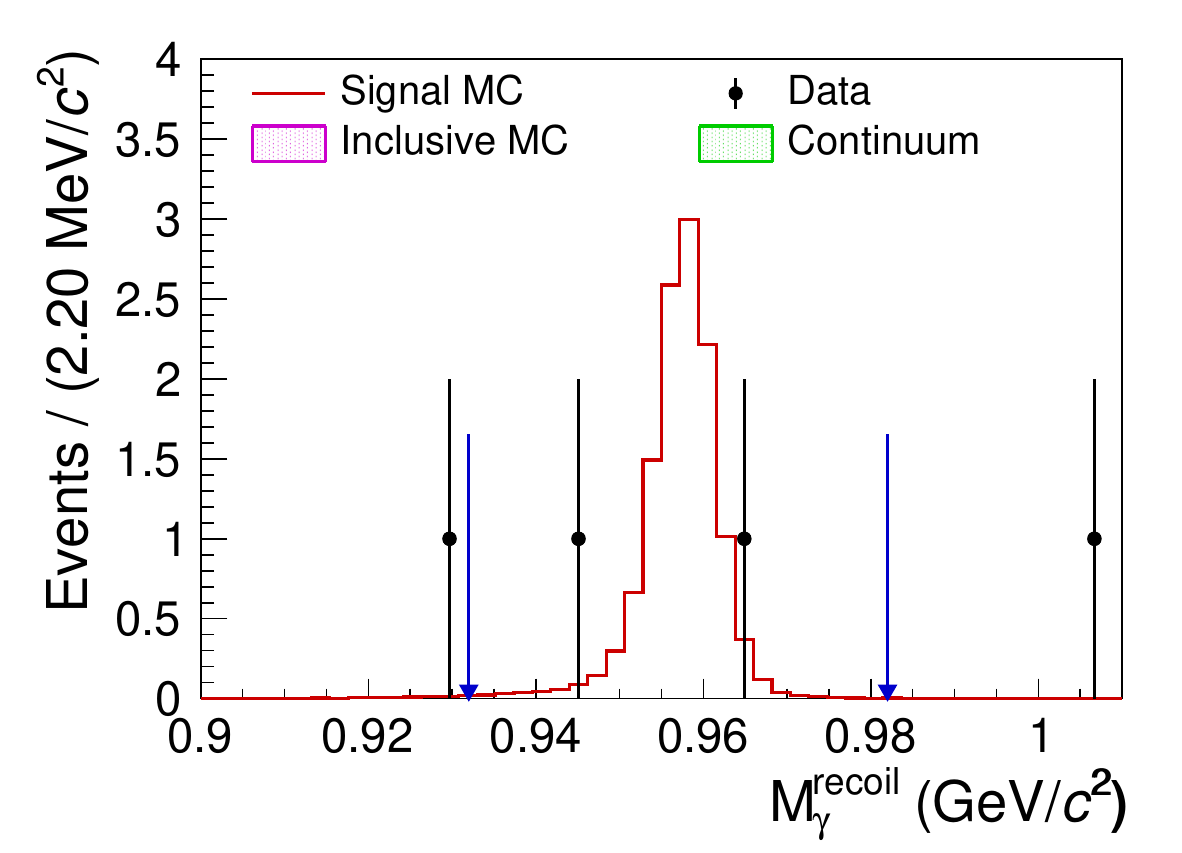}
    \caption{The $M_\gamma^{\rm recoil}$ distribution of the accepted candidate events. The red histogram represents signal MC sample and the black dots with error bars represent the data sample. No event from inclusive MC sample or continuum samples fall within the plotted range. The blue arrows indicate the signal region.}
    \label{fg:3}
\end{figure}

Finally, events with the photon recoil mass  $M_{\gamma}^{\rm recoil} = \sqrt{(p_{J/\psi}-p_\gamma)^2}$ (where $p$ denotes the particle four-momentum) in the signal range $(0.932,0.982)~{\rm GeV}/c^2$ are considered as the signal candidates. The signal mass window corresponds to five times the mass resolution around the nominal $\eta'$ mass. As shown in Figure~\ref{fg:3}, after applying all the event selection criteria mentioned above, two $\eta'\to e\mu$ candidates are observed in the signal region. No events remain in the inclusive MC sample and continuum samples. The signal selection efficiency is $(17.93\pm0.06)\%$.

%%====================================================

\section{Background study}
\label{sec:bkg}

The $J/\psi$ decays and continuum processes are two potential background sources for the analysis.
The background yield from $J/\psi$ decays is estimated by analyzing the inclusive $J/\psi$ MC sample.
The number of events surviving in the signal region is zero.

The contribution of continuum processes is estimated using the data samples collected at $\sqrt{s}=$ $3.080~{\rm GeV}$, $3.650~{\rm GeV}$, $3.682~{\rm GeV}$ and $3.773~{\rm GeV}$. The cross section of the continuum process is assumed to have an inverse-square of $\sqrt{s}$-dependence, so the normalized background level from the continuum processes can be obtained by multiplying the number of surviving events from each data set by the scale factor $(f_i)$ and the weight $(w_i)$, which are defined in Section~\ref{sec:bes}.
No events survive in the signal region, so the normalized continuum background contribution is estimated to be zero.

By summing the background contributions from $J/\psi$ decay and continuum background, we obtain a total background yield of zero.

%%====================================================

\section{Systematic uncertainties}
\label{sec:error}

The systematic uncertainties mainly arise from the tracking and PID efficiencies of the electron and the muon, other signal selection criteria, the number of $J/\psi$ events in the data ($N_{J/\psi}$), the BF of $J/\psi\to\gamma\eta'$ process ($\mathcal{B}(J/\psi\to\gamma\eta')$), and the statistical uncertainty of the signal MC efficiency. The systematic uncertainties from these sources are studied as follows and summarized in Table~\ref{tb:5.1}. The formula for calculating $\mathcal{B}(\eta'\to e\mu)$ is 
\begin{equation}\label{eq:5.1}
    \mathcal{B}(\eta'\to e\mu) = \dfrac{N_{\rm sig}}{N_{\eta'}\cdot\varepsilon},\quad N_{\eta'} = N_{J/\psi}\cdot\mathcal{B}(J/\psi\to\gamma\eta').
\end{equation}
where $N_{\rm sig}$ is the number of signal events. The uncertainties of $N_{J/\psi}$ and $\mathcal{B}(J/\psi\to\gamma\eta')$ are then combined in quadrature to obtain the 
$N_{\eta'}$ uncertainty $\Delta_{N_{\eta'}}$. The rest of uncertainty terms all contribute to the signal efficiency $(\varepsilon)$, and are combined in quadrature to derive its total uncertainty $\Delta_{\varepsilon}$.
The processes $e^+e^-\to \gamma e^+e^-$ and $e^+e^-\to\gamma\mu\mu$ are used as 
the control samples to estimate the uncertainties 
caused by $\emph{Tracking}$, \emph{PID},
\emph{Photon selection}, and \emph{4C kinematic fit} as listed in the following items.

\begin{table}[!ht]
\centering
\caption{Summary of systematic uncertainties from all considered sources. The first five sources contribute to the uncertainty of signal efficiency, while the last two sources contribute to the uncertainty of the number of $\eta'$.}
\label{tb:5.1}
    \begin{tabular}{clc}
    \toprule 
    Source & \multicolumn{1}{c}{\quad Uncertainty/\%} & \multicolumn{1}{c}{Combined/\%}\\
    \midrule
    Tracking & \qquad 2.0~\cite{bib:5-1, bib:5-1.1}& \multirow{5}{*}{$\Delta_{\varepsilon}=7.5$} \\ %\cite{bib:5-1,bib:5-1.1}
    PID & \qquad 6.9 \\
    Photon selection    & \qquad0.5 \\ % \cite{bib:5-2}
    4C kinematic fit    & \qquad1.7 \\
    Statistics of MC sample & \qquad0.4 \\
    \cmidrule(lr){1-3}
    Number of $J/\psi$ events & \qquad0.5~\cite{bib:3-1} & \multirow{2}{*}{$\Delta_{N_{\eta'}}=1.3$}\\% 0.45+0.56*224/8774=0.46
    BF of $J/\psi\to\gamma\eta'$ process & \qquad1.2~\cite{pdg} \\
    \bottomrule
    \end{tabular}
\end{table}

\begin{itemize}
\item \emph{Tracking.}
The relative efficiency difference between the data and MC sample of the control sample is taken as the systematic uncertainty: 1.0\% for each charged track conservatively~\cite{bib:5-1, bib:5-1.1}.

\item \emph{PID.}
The combined systematic uncertainty of 
the electron and muon PID is 6.9\%.

\item \emph{Photon selection.}
The control sample of $e^+e^-\to\gamma\mu^+\mu^-$ process is used to study the photon selection efficiency. The relative efficiency difference between data and MC samples are observed to be up to $0.5\%$.

\item \emph{4C kinematic fit.}
The relative efficiency differences between the data and MC sample for the control samples, after applying $\chi_{\rm 4C}^2$ requirement, are $0.3\%$ for $\gamma e^+e^-$ sample and $1.7\%$ for $\gamma\mu^+\mu^-$. The larger value $1.7\%$ is taken as the systematic uncertainty.

\item \emph{Statistics of signal MC sample.}
The statistical uncertainty of the signal efficiency is estimated from the standard deviation of the binomial-distributed signal yield. The resulting uncertainty is 0.4\%.

\item \emph{Total number of ${J/\psi}$ events.}
The total number of $J/\psi$ events, $(8998 \pm 40) \times 10^6$, is taken from the BESIII measurement~\cite{bib:3-1}, and its uncertainty is 0.5\%. 

\item $\mathcal{B}(J/\psi\to\gamma\eta')$. The uncertainty of $\mathcal{B}(J/\psi\to\gamma\eta')$ is 1.2\% according to the PDG~\cite{pdg}.
\end{itemize}

%%====================================================

\section{Result}
\label{sec:result}

We observe two candidate signal events in the
$\eta^{\prime}$ signal region, with zero expected background events. 
Nevertheless, no significant excess over background is observed, as the statistics of the inclusive MC sample is similar to that of the experimental data.
An UL on $\mathcal{B}(\eta'\to e\mu)$ is estimated with the profile likelihood method~\cite{bib:6-1}. The likelihood function is defined as
\begin{equation}\label{eq:6.1}
\begin{split}
\mathcal{L}(\mathcal{B},\varepsilon,N_{\eta'},{N}_{\rm bkg1},{N}_{\rm bkg2})
&= \mathcal{P}\qty(\left.N_{\rm obs}~\right|~N_{\eta'}\cdot\mathcal{B}\cdot\varepsilon+N_{\rm bkg1}+N_{\rm bkg2}) \\
&~~\cdot\mathcal{P}\qty(\left.N_{\rm bkg1}^{\rm inc}~\right|~{N}_{\rm bkg1}) \\
&~~\cdot\prod_i\mathcal{P}\qty(\left.N_{\rm bkg2}^i~\right|~{N}_{\rm bkg2}/f_i) \\
&~~\cdot\mathcal{G}\qty(\varepsilon~\left|~\varepsilon_{\rm MC},\sigma_{\varepsilon}\right.) \\
&~~\cdot\mathcal{G}\qty(N_{\eta'}~\left|~N_{\eta'}^{\rm data},\sigma_{N_{\eta'}}\right.),
\end{split}
\end{equation}
where the number of observed events ($N_{\rm obs}$) follows a Poisson distribution $(\mathcal{P})$ with an expected number given by the expected signal events $(N_{\eta'}\cdot\mathcal{B}\cdot\varepsilon)$ and two types of background events $(N_{\rm bkg1}+N_{\rm bkg2})$. The numbers of different background components follow Poisson distributions. The number of background events observed in $J/\psi$ inclusive MC sample and the continuum data samples are $N_{\rm bkg1}^{\rm inc}$ and $N_{\rm bkg2}^i$, respectively. The scale factors of the continuum data samples are denoted as $f_i$. The signal detection efficiency follows a Gaussian distribution $(\mathcal{G})$ with the values of mean and standard deviation given by MC-determined nominal efficiency ${\varepsilon}_{\rm MC}$ and its absolute uncertainty $\sigma_\varepsilon = {\varepsilon}_{\rm MC}\cdot\Delta_{\varepsilon}$, respectively. Similarly, for the $\eta'$ number, the mean and standard deviation are $N_{\eta'}^{\rm data}=N_{J/\psi}^{\rm data}\cdot\mathcal{B}(J/\psi\to\gamma\eta')$ and $\sigma_{N_{\eta'}}=N_{\eta'}^{\rm data}\cdot\Delta_{N_{\eta'}}$, respectively. Here, $N_{J/\psi}^{\rm data}$ is the number of $J/\psi$ events in data and the $\mathcal{B}(J/\psi\to\gamma\eta')$ value is taken from the PDG~\cite{pdg}. All the fixed parameters are shown in Table~\ref{tb:6.1}.
	
\begin{table}[!ht]
\centering
\renewcommand\arraystretch{1.3}
\caption{The values of the fixed parameters in the likelihood function in Formula~\eqref{eq:6.1}.}
\label{tb:6.1}
\begin{tabular}{p{8em}<{\centering}p{8em}<{\centering}|p{8em}<{\centering}p{8em}<{\centering}}
\hline
Parameter & Value & Parameter & Value \\
\hline
    $N_{\rm obs}$               & $2$   & $N_{\eta'}^{\rm data},\ \sigma_{N_{\eta'}}/N_{\eta'}^{\rm data}$ & $4.75\times10^{7},\ 1.3\%$ \\
    ${N}_{\rm bkg1}^{\rm inc}$  & $0$   & ${\varepsilon}_{\rm MC},\ \sigma_{\varepsilon}/{\varepsilon}_{\rm MC}$ & $17.93\%,\ 7.5\%$ \\
    ${N}_{\rm bkg2}^{3.080}$    & $0$   & $f_{3.080}$ & $16.42$   \\
    ${N}_{\rm bkg2}^{3.650}$    & $0$   & $f_{3.650}$ & $8.44$    \\
    ${N}_{\rm bkg2}^{3.682}$    & $0$   & $f_{3.682}$ & $9.66$    \\
    ${N}_{\rm bkg2}^{3.773}$    & $0$   & $f_{3.773}$ & $0.52$    \\
\hline
\end{tabular}
\end{table}

\begin{figure}[!ht]
\centering
\includegraphics[width=0.49\textwidth]{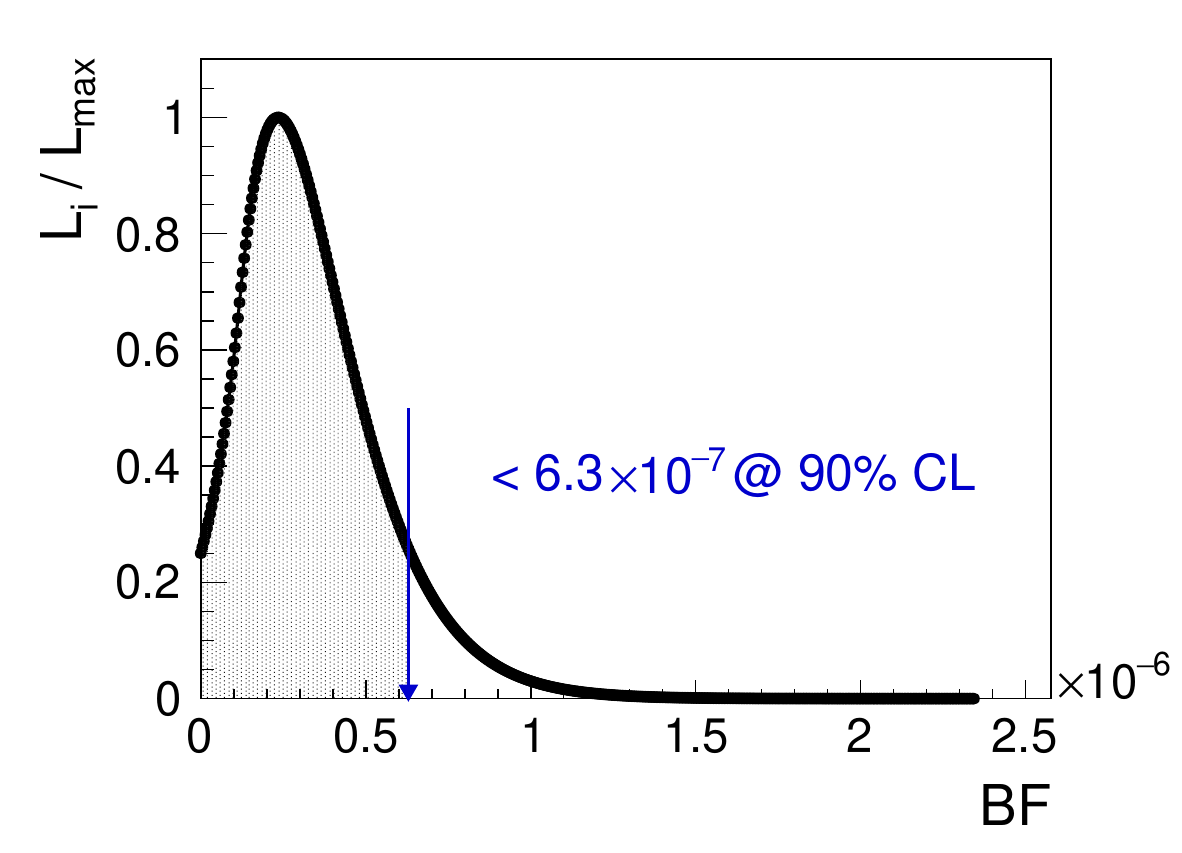}
\caption{The normalized likelihood distribution as a function of $\mathcal{B}(\eta'\to e\mu)$. The blue arrow indicates the UL of BF at the 90\% CL.}
\label{fg:6.1}
\end{figure}

For each assumed BF value, the likelihood function in Formula~\eqref{eq:6.1} is maximized with respect to all remaining nuisance parameters, yielding the corresponding local maximum likelihood. By scanning over a range of BF values, the likelihood distribution, denoted by $\mathcal{L}_i$, is obtained. The normalized profile likelihood, defined as $\mathcal{L}_i/\mathcal{L}_{\rm max}$, where $\mathcal{L}_{\rm max}$ is the global maximum likelihood, is shown as a function of $\mathcal{B}$ in Figure~\ref{fg:6.1}. By integrating the likelihood distribution in the physical region until 90\% of the total area is accumulated, the UL on the BF of $\eta'\to e\mu$ is determined to be $6.3\times10^{-7}$ at the 90\% CL. This result improves upon the previous best limit~\cite{bib:1-10} by nearly three orders of magnitude.

According to Reference~\cite{bib:1-7}, for pseudoscalar mesons, the BFs are directly proportional to the square of the Wilson coefficients over $\Lambda^2$ with
\begin{equation}\label{eq:6.2}
    \mathcal{B}({\rm P}\to\ell_1\ell_2) \propto \abs{C/\Lambda^2}^2,
\end{equation}
where $C$ is the Wilson coefficient, which can be $C_{\rm AL/AR}^{q\ell_1\ell_2}$, $C_{\rm PL/PR}^{q\ell_1\ell_2}$ or $C_{\rm \tilde{G}L/\tilde{G}R}^{\ell_1\ell_2}$. Since the Wilson coefficients $C_{\rm AL/AR}^{q\ell_1\ell_2}$, $C_{\rm PL/PR}^{q\ell_1\ell_2}$ are associated with quark flavour, the quark components $u\bar{u}+d\bar{d}$ and $s\bar{s}$ of $\eta'$ impose constraints on different coefficients. The parameter $\Lambda$ is the new physics energy scale. When setting a constraint on a single Wilson coefficient, the assumption of single operator dominance is made, meaning contributions from other Wilson coefficients are neglected. As a result, the constraints on Wilson coefficients are improved by $1\sim2$ orders of magnitude, as shown in Table~\ref{tb:6-2}. 

\begin{table}[!ht]
\centering
\caption{Comparison of the ULs on the Wilson coefficients from $\eta'$ meson decays obtained in this analysis and from previous analysis. The ULs of $\mathcal{B}(\eta'\to e\mu)$ are provided at 90\% CL.}
\label{tb:6-2}
\renewcommand{\arraystretch}{1.6}
\begin{tabular}{ccllll}
\toprule
Wilson coefficients & \multirow{2}{*}{Final states} & \multicolumn{2}{c}{Previous result~\cite{bib:1-10}} & \multicolumn{2}{c}{This analysis} \\
\cmidrule(lr){3-6}
over $\Lambda^2$ & & \multicolumn{2}{c}{$\mathcal{B}(\eta'\to e\mu) < 4.7\times10^{-4}$} & \multicolumn{2}{c}{$\mathcal{B}(\eta'\to e\mu) < 6.3\times10^{-7}$} \\
\cmidrule(lr){3-6}
$({\rm GeV}^{-2})$& $\ell_1\ell_2$ & \multicolumn{1}{c}{$q=u,d$} & \multicolumn{1}{c}{$q=s$} & \multicolumn{1}{c}{$q=u,d$} & \multicolumn{1}{c}{$q=s$} \\ 
\midrule
\multirow{1}{*}{$\abs{C_{\rm AL/AR}^{q\ell_1\ell_2}/\Lambda^2}$} & $e\mu$ & $2.1\times10^{-1}$ & $1.9\times10^{-1}$ & $7.4\times10^{-3}$ & $6.8\times10^{-3}$ \\
\multirow{1}{*}{$\abs{C_{\rm PL/PR}^{q\ell_1\ell_2}/\Lambda^2}$} & $e\mu$ & $3.9\times10^{4}$ & $3.6\times10^{4}$ & $1.4\times10^{3}$ & $1.3\times10^{3}$ \\
\multirow{1}{*}{$\abs{C_{\rm \tilde{G}L/\tilde{G}R}^{\ell_1\ell_2}/\Lambda^2}$} & $e\mu$ & \multicolumn{2}{c}{$5.0\times10^{3}$} & \multicolumn{2}{c}{$1.8\times10^{2}$}  \\
\bottomrule
\end{tabular}
\end{table}

%%====================================================

\section{Summary}
\label{sec:sum}

Using the world's largest $J/\psi$ data sample collected with the BESIII detector, a search for the charged lepton flavour violating decay $\eta'\to e\mu$ has been performed. Based on $(8998\pm40)\times 10^{6}~J/\psi$ events~\cite{bib:3-1}, no statistically significant excess above the expected background is observed. The UL on $\mathcal{B}(\eta'\to e\mu)$ is set to $6.3\times10^{-7}$ at 90\% CL, improving the previous limit of $\eta'\to e\mu$~\cite{bib:1-10} by nearly three orders of magnitude, the most stringent constraint on this decay channel to date. Consequently, the constraints on Wilson coefficients $|C_{\rm AL/AR}^{qe\mu}/\Lambda^2|$, $|C_{\rm PL/PR}^{qe\mu}/\Lambda^2|$ and $|C_{\rm \tilde{G}L/\tilde{G}R}^{e\mu}/\Lambda^2|$ (where $q=u,d$) derived by this decay channel are tightened by more than one order of magnitude. These results provide complementary constraints on CLFV interactions in the pseudoscalar sector and contribute to the broader experimental program searching for physics beyond the Standard Model.

%%====================================================

% \appendix
% \section{Some title}
% Please always give a title also for appendices.

%%====================================================

\acknowledgments

%% Saved at => 2026-03-02
% \textbf{Acknowledgement}

The BESIII Collaboration thanks the staff of BEPCII (https://cstr.cn/31109.02.BEPC) and the IHEP computing center for their strong support. This work is supported in part by National Key R\&D Program of China under Contracts Nos. 2023YFA1606000, 2023YFA1606704, 2025YFA1613900; National Natural Science Foundation of China (NSFC) under Contracts Nos. 11635010, 11935015, 11935016, 11935018, 12025502, 12035009, 12035013, 12061131003, 12192260, 12192261, 12192262, 12192263, 12192264, 12192265, 12221005, 12225509, 12235017, 12342502, 12361141819, 12535005; the Chinese Academy of Sciences (CAS) Large-Scale Scientific Facility Program; the Strategic Priority Research Program of Chinese Academy of Sciences under Contract No. XDA0480600; CAS under Contract No. YSBR-101; 100 Talents Program of CAS; The Institute of Nuclear and Particle Physics (INPAC) and Shanghai Key Laboratory for Particle Physics and Cosmology; ERC under Contract No. 758462; German Research Foundation DFG under Contract No. FOR5327; Istituto Nazionale di Fisica Nucleare, Italy; Knut and Alice Wallenberg Foundation under Contracts Nos. 2021.0174, 2021.0299, 2023.0315; Ministry of Development of Turkey under Contract No. DPT2006K-120470; National Science and Technology fund of Mongolia; Polish National Science Centre under Contract No. 2024/53/B/ST2/00975; STFC (United Kingdom); Swedish Research Council under Contract No. 2019.04595; U. S. Department of Energy under Contract No. DE-FG02-05ER41374

\bibliographystyle{JHEP}
\bibliography{biblio.bib}

\section*{The BESIII Collaboration}\label{author}
%% Saved at => 2026-03-02
\noindent
M.~Ablikim$^{1}$\BESIIIorcid{0000-0002-3935-619X},
M.~N.~Achasov$^{4,c}$\BESIIIorcid{0000-0002-9400-8622},
P.~Adlarson$^{83}$\BESIIIorcid{0000-0001-6280-3851},
X.~C.~Ai$^{89}$\BESIIIorcid{0000-0003-3856-2415},
C.~S.~Akondi$^{31A,31B}$\BESIIIorcid{0000-0001-6303-5217},
R.~Aliberti$^{39}$\BESIIIorcid{0000-0003-3500-4012},
A.~Amoroso$^{82A,82C}$\BESIIIorcid{0000-0002-3095-8610},
Q.~An$^{78,65,\dagger}$,
Y.~H.~An$^{89}$\BESIIIorcid{0009-0008-3419-0849},
M.~S.~Anderson$^{39}$\BESIIIorcid{0009-0008-1550-2632}
% ...
\\
% \vspace{0.2cm}
% (BESIII Collaboration)\\
{\it\\
$^{1}$ Institute of High Energy Physics, Beijing 100049, People's Republic of China\\
% ...
$^{4}$ Budker Institute of Nuclear Physics SB RAS (BINP), Novosibirsk 630090, Russia\\
% ...
$^{31}$ INFN Sezione di Ferrara, (A)INFN Sezione di Ferrara, I-44122, Ferrara, Italy; (B)University of Ferrara, I-44122, Ferrara, Italy\\
% ...
$^{39}$ Johannes Gutenberg University of Mainz, Johann-Joachim-Becher-Weg 45, D-55099 Mainz, Germany\\
% ...
$^{65}$ State Key Laboratory of Particle Detection and Electronics, Beijing 100049, Hefei 230026, People's Republic of China\\
% ...
$^{78}$ University of Science and Technology of China, Hefei 230026, People's Republic of China\\
% ...
$^{82}$ University of Turin and INFN, (A)University of Turin, I-10125, Turin, Italy; (B)University of Eastern Piedmont, I-15121, Alessandria, Italy; (C)INFN, I-10125, Turin, Italy\\
$^{83}$ Uppsala University, Box 516, SE-75120 Uppsala, Sweden\\
% ...
$^{89}$ Zhengzhou University, Zhengzhou 450001, People's Republic of China\\

\noindent
$^{\dagger}$ Deceased\\
% ...
$^{c}$ Also at the Novosibirsk State University, Novosibirsk, 630090, Russia\\
% ...
}
%% ends here %%

\end{document}